\documentclass[12pt]{iopart}
\pdfoutput=1

\usepackage{graphicx,epsfig,floatflt}
\usepackage{amssymb}
\usepackage{iopams}
\usepackage{float}
\usepackage{multirow}
\usepackage{url}

\begin{document}
\input {epsf}
\def\plotone#1{\centering \leavevmode
\epsfxsize= 0.4\columnwidth \epsfbox{#1}}
\def\plottwo#1#2{\centering \leavevmode
\epsfxsize=.43\columnwidth \epsfbox{#1} \hfil
\epsfxsize=.43\columnwidth \epsfbox{#2}}
\def\plotfiddle#1#2#3#4#5#6#7{\centering \leavevmode
\vbox to#2{\rule{0pt}{#2}}
\includegraphics{#1}}
\newdimen\hhsize\hhsize=.5\hsize
\font\lloyd=cmr8 

\def\Journal#1#2#3#4{{#1} {\bf #2}, #3 (#4)} 
\def\NCA{\em Nuovo Cimento} \def\NIM{\em Nucl. Instrum. Methods}
\def\NIMA{{\em Nucl. Instrum. Methods} A} \def\NPB{{\em Nucl. Phys.} B}
\def\PLB{{\em Phys. Lett.}  B} \def\PRL{{\em Phys. Rev. Lett.}}
\def\PRD{{\em Phys. Rev.} D} \def\ZPC{{\em Z. Phys.} C}

\def\today{\ifcase\month\or
 January\or February\or March\or April\or May\or June\or
 July\or August\or September\or October\or November\or December\fi
 \space\number\day, \number\year}

\title{Constraints on Tensor to Scalar Ratio using WKB
Approximation}
\author{Aiswarya. A$^{a,b}$ and Minu Joy$^{a}$ }
\address{$^a$ 
Dept. of Physics, Alphonsa College, Pala 686574, India}
\address{$^b$ 
School of Pure and Applied Physics, Mahatma Gandhi University, Kottayam 686560, India}

\begin{abstract}
\small{ Using Wenzel-Kramers-Brillouin (WKB) approximation the scalar and tensor power spectra are obtained. Scale invariant spectra are obtained 
and the spectral indices come very close to the observed data from WMAP and Planck experiments. The advantage of this method is that, it is valid even when
slow-roll approximation fails. Constraints on the tensor to scalar ratio(\textit{r}) is also studied with the WKB Approximation. We use the Power law inflation 
as the base model as it allows comparison with exact results. }
\end{abstract}
\maketitle

\section{Introduction}
\label{sec:intro} 
Inflationary cosmology has come a long way, starting from being the theory to explain the shortcomings of Big Bang model 
\cite{starbsky, guth, linde} to the most accepted model with 
the current observations. Success of $NASA's$ first CMB satellite COBE (Cosmic Background Explorer satellite), paved way for future projects in CMB. 
These results gave an accurate thermal spectrum of CMB, the first detection of primordial CMB fluctuations with normalization at 10$^{-5}$.
It also gave cosmic variance limit  of large angular scale power spectrum \cite{cobe}.
Among the CMB missions which followed COBE, BICEP mission was designed to measure the B-mode polarization
at degree angular
scales and can be considered as the  first mission to constrain the tensor-to-scalar ratio $r$ directly from CMB B-mode polarization. 
The mission gave the $r$ constraints as 
\begin{equation*}
r=0.02^{+0.31}_{-0.26} \quad or \quad r<0.72 \quad at \quad 95\%\quad C.L.
\end{equation*}

The direct  measurement  of $r$ from the BB spectrum  has advantage, as the  measurements of TT are limited by cosmic variance at large angular scales
and the $r$ constraints from TT have strong  parameter degeneracies such as with the 
scalar spectral index $n_s$, while the BB amplitude depends primarily on \textit{r}
\cite{bicep}.

The SPT-SZ survey helped to remove the degeneracy between $n_s$ and $r$ by measurements of the power spectrum from angular scales corresponding
 to multipole range 650 $<l<$ 3000, that is from the third acoustic peak through the CMB damping tail. The constrain on scalar spectral index tightens 
to $n_s<1.0$ at $3.9\sigma$ by using SPT and WMAP7 data \cite{spt}. 
The B and E mode  polarization data gives constraints on the recombination (last-scattering) and reionization epochs. The cross-correlation between 
temperature anisotropies and polarization signal is a  consistency check for inflationary cosmology and helps in breaking degeneracies among the cosmological 
parameters \cite{cmbpol}.

Analysis of the combined BICEP2 and Keck Array data in combination with the 2015 Planck data put an upper limit $r_{0.05} < 0.12 $ at 95$\%$ 
confidence \cite{comb}. 
2018 release of the Planck temperature, polarization and lensing
data determine the spectral index of scalar perturbations to be 
\begin{equation*}
n_s= 0.965\pm0.004
\end{equation*}
at $68\%$ CL and find no evidence for a scale dependence
of $n_s$, either as a running or as a running of the running.
The latest Planck $95\%$ CL upper limit on the tensor-to-scalar ratio,
$r_{0.002} < 0.10 $, is further tightened by combining with the BICEP2/Keck
Array BK14 data to obtain $r_{0.002} < 0.064 $ \cite{planck2018t, planck2018}.

In the present work we study the constraints on the tensor-to-scalar ratio \textit{r} using an alternative approximation technique, instead of 
slow-roll approximation, with the Power law inflationary model. We first enumerate the  impact of \textit{r} on different inflationary models and cosmological parameters, 
followed by a brief review on some alternative approximation techniques. Scalar and tensor power spectra 
using the WKB approximation technique are obtained. Hence we try to constrain tensor to scalar ratio with WKB approximation and check the consistency 
with latest Planck observations.

\section{Tensor to scalar ratio}
The study of tensor to scalar ratio \textit{r} gives insight into large working of the universe :- 
\begin{itemize}
\item
Primordial Gravitational waves - The tensor perturbations in the FLRW background metric during inflation are called  the primordial gravitational waves and CMB is 
the only observation to validate it. The CMB sky is measured with two observables, the temperature anisotropy and the polarisation. Primordial gravitational 
waves imprints very slight variations on these variables and are sensitive to \textit{r} which is defined as \cite{cmbpol}
\begin{equation*}
r=\frac{\mathcal{P}_t}{\mathcal{P}_s}
\end{equation*}
It contributes to the temperature anisotropies as  a local quadruple since the gravitational wave metric is traceless. The primordial tensor power spectrum \cite{star}
 is scale
 invariant and the shear is impulsive at horizon entry \cite{wang}. \\

\item Energy scale of universe -  \textit{r} depends on the time-evolution of the inflaton field and the relation to inflaton potential \cite{liddle} is given as
\begin{equation*}
r=\frac{8}{M_{Pl}^2}\bigg(\frac{\dot\phi}{H}\bigg)^2
\end{equation*}
\begin{equation*}
V^{1/4}=1.06\times10^{16} \bigg(\frac{r_{\phi}}{0.01}\bigg)^{1/4} GeV
\end{equation*} where $r_{\phi}$ is the tensor to scalar ratio in CMB scales \cite{cmbpol}.
\item Consistency relation -  \textit{r} gives a test for consistency for single-field slow roll inflation as
\begin{equation*}
r=-8n_{t}sin^2\Delta
\end{equation*}
where $sin^2\Delta$ parameterizes the ratio between the adiabatic power spectrum at horizon-exit during inflation and the observed power spectrum giving information about presence of multiple fields. For non-slow roll inflation, with a non-trivial $c_s$, it is given as
\begin{equation*}
r=-n_{t}c_s
\end{equation*}
where it is seen that the non-slow roll evolution of the inflaton field is driven by a non-canonical kinetic term.
\item Inflaton Field Excursion -  The inflaton field $\Delta \phi$ is related to  \textit{r} giving the effective number of  e-folds as
\begin{equation*}
N_{eff}=\int_{0}^{N_{\star}}dN\bigg(\frac{r(N)}{r_{\star}}\bigg)^{\frac{1}{2}}
\end{equation*} 
where  $N_{\star}$ is the number of e-folds between the end of inflation and the horizon exit of the CMB pivot scale. By determining,
 if the inflaton-field excursion was super-planckian or not we can probe aspects of scalar field space and the UV completion of gravity.

\item Inflationary models -  The shape of the inflaton potential is controlled by the slow-roll parameters and hence measurements of 
\textit{r}, $n_s$, $n_t$, $\alpha_{s}$, $\alpha_t$ help in constraining different inflationary models and also removes the degeneracies between the slow-roll parameters.
\end{itemize}

\section{Alternatives for Standard Slow-roll approximation}
There are many models proposed which give the similar observations, which doesn't explicitly consider slow roll approximation for inflation  but gives very 
similar results without the approximation \cite{martin}. For example,

\subsection{Hamilton-Jacobi formulation of inflation}
Hamilton-Jacobi formulation \cite{mus, bond} is a method to rewrite the equations of motion. It considers the scalar inflaton field  itself to be 
time-varying but it breaks down during the oscillatory epoch that ends inflation \cite{liddlelyth, minu2}.

We know the  equations of motion for a spatially flat universe as

\begin{equation}
H^2=\frac{1}{3M_{pl}^{2}}(V(\phi)+\frac{1}{2}\dot{\phi^{2}})
\end{equation}
and 
\begin{equation}
\ddot{\phi}+3H\dot{\phi}=-\frac{dV}{d\phi} ,
\end{equation} where $V(\phi)$ is the potential of the scalar field.
By differentiating the first equation of motion with respect to time and by substituting in the second equation of motion, we get
\begin{equation}
2\dot{H}=-\frac{\dot{\phi}^{2}}{M_{pl}^{2}}
\end{equation}
Dividing both sides by $\dot{\phi}$ gives,
\begin{equation}
\dot{\phi}=-2M_{pl}^{2}H'(\phi)
\end{equation}
which gives the Freidmann-equations in the first order form,
\begin{equation}
H'(\phi)^{2}-\frac{3}{2M_{pl}^{2}}H^{2}(\phi)=-\frac{1}{2M_{pl}^{4}}V(\phi) .
\end{equation}
When $H(\phi)$ is specified it gives the corresponding potential and all the other inflationary solutions can be derived from it.
For example,
\begin{equation}
H(\phi)\propto exp\Big(-\sqrt{\frac{1}{2p}}\frac{\phi}{M_{pl}}\Big)
\end{equation}
gives the Power-law inflation.
The slow-roll parameters  can be written as,
\begin{equation}
\epsilon_{H}=3\frac{\dot{\phi}^2}{2}/\Big(V+\frac{\dot{\phi}^2}{2}\Big)=-\frac{dlnH}{dlna}
\end{equation}
\begin{equation}
\eta_{H}=-\frac{\ddot{\phi}}{H\dot{\phi}}=-\frac{dln\dot{\phi}}{dlna}
\end{equation}
and inflation is precisely given as,
\begin{equation}
\ddot{a}>0 \Longleftrightarrow\epsilon_{H}<1 .
\end{equation}

\subsection{Perturbed Power law inflation}
With the standard Power law model the scale factor is given as, $a(t)=t^p$
where $p>1$ and the slow-roll parameters are given as $\epsilon=-\eta=\frac{1}{p}$. 
The Power-law also states that \cite{lucchin} - \cite{ts_thesis}, 
\begin{itemize}
\item[(a)]Expansion is uniformly accelerated \textit{ie }$\epsilon=constant$
\item[(b)]The perturbations in the inflaton field is similar to that for a massless scalar field $m^2_{eff}=0$.
\end{itemize}
The power spectrum is given as $\mathcal{P}_{k}=Ak^{n_s -1}$ and spectral indexes are $n_s$ and $n_t=n_s-1$ which are constant values. 

The perturbed Power law \cite{minu1, ppl2} can be given as a slight deviation from the scale invariant spectra. Here, the inflaton potential is estimated from the `reduced Hamilton -Jacobi equation'. 
The corresponding slow-roll parameters are  given as,
$\epsilon=\frac{M_{pl}^2}{4\pi}\Big(\frac{H'}{H}\Big)^2$ and $\eta=-\frac{M_{pl}^2}{4\pi}\Big(\frac{H''}{H}\Big)$. The spectral indices are then,
\begin{equation}
n_t=n_s-1=\frac{-2\epsilon}{1-\epsilon}
\end{equation}where $\epsilon=(\frac{\partial ln H}{\partial \phi})^2$.

The Perturbed power-law is taken for small deviations from the uniform acceleration in terms of $\frac{\partial^2 lnH}{\partial\phi^2}$. The scalar perturbation spectrum 
is then obtained as,
\begin{equation}
\mathcal{P}_{s}(k)=\frac{A(\mu_T , \mu_S)}{2\pi}\Big(\frac{H}{2\pi}\Big)^2 \frac{1}{\epsilon(\eta)}
\end{equation} and the corresponding tensor power-spectrum is,
\begin{equation}
\mathcal{P}_t(k)=\frac{8A(\mu_T , \mu_S)}{2\pi} \Big(\frac{H}{2\pi}\Big)^2
\end{equation}
where $A(x,y)=\frac{4^y\Gamma^2(y)}{(x-1/2)^{2y-1}}$, $\Gamma(y)$ is the Euler gamma function where $\mu_T\rightarrow \mu $ and \\ 
$\mu_s$ = $\Big(\mu^2 - (\mu-1/2)^2\frac{m^2_{eff}}{H^2}\Big)^{1/4}$.

\section{WKB approximation}
The WKB approximation is a method used to get approximate solutions to linear differential equations. Here it is applied on the Radial wave equations.
 It was proved that the Balmer formula for Hydrogen atom could be derived from correct application of the WKB approximation. Hence the method was used for 
cosmological perturbations which had an analogy with the Hydrogen atom formalism. The use of this method also gives the advantage that it can be used for
 both subhorizon and the superhorizon scales and it is also an alternate method than the standard slow-roll approximation. 

The non-singular solutions with initial dense de-sitter phase, first mentioned by Starobinsky gave 
the tensor perturbation formulation \cite{star}. Mukhanov and Chibisov proposed that the spectrum of quantum fluctuation in the non-singular solution could evolve
into the present universe and derived the 
scalar perturbation equation \cite{mukh}, followed by works of Hawking \cite{hawk} and Starobinsky \cite{star2}.
These scalar and tensor perturbations formulations are reduced to a single variable \cite{guthpi} denoted by $\mu_S$ or $\mu_T$, and the equation is similar to 
the Schr$\ddot{o}$dinger equation \cite{Schro}.
\begin{equation}
\frac{d^2}{d\eta^2}\mu +\big(k^2-U(\eta)\big)\mu = \frac{d^2\mu}{d\eta^2}+\omega^2(\eta)\mu =0
\end{equation}where $U(\eta)$ is the potential. Then the corresponding power spectra and the spectral indices are given as,
\begin{equation}
\mathcal{P}_s=\frac{k^3}{8\pi^2}\Big|\frac{\mu_s}{z_s}\Big|^2, \quad \mathcal{P}_T=\frac{2k^3}{\pi^2}\Big|\frac{\mu_T}{z_T}\Big|^2
\end{equation}
where $z_s=a\sqrt{\frac{-aa''}{a'}}$ and $z_T=a$. The spectral indices being,
\begin{equation}
n_s-1=\Big|\frac{dln \mathcal{P}_s}{dlnk}\Big|_{k=k_*},\quad n_T=\Big|\frac{dln\mathcal{P}_T}{dlnk}\Big|_{k=k_*} .
\end{equation}

The WKB spectra is taken as the solution for the equation,
\begin{equation}
\mu''_{WKB}(k,\eta)+[\omega^2(k,\eta)-Q(k,\eta)]\mu_{WKB}=0
\end{equation} where 
\begin{equation}
Q(k,\eta)=\frac{3}{4}\frac{(\omega')^2}{\omega}-\frac{\omega''}{2\omega} .
\end{equation}
The mode function $\mu_{WKB}$ is an approximation to the actual mode function in the limit,
\begin{equation}
\Big|\frac{Q}{\omega^2}\Big|<1
\end{equation}

For application of the WKB approximation, the variables are transformed to 
\begin{equation}
x=ln\Big(\frac{Ha}{k}\Big),\quad u=(1-\epsilon_1)^{\frac{1}{2}}e^{\frac{x}{2}}\mu
\end{equation}

which in turn gives the Power spectra to be,
\begin{equation}
\mathcal{P}_s=\frac{H^2}{\pi\epsilon_1M_{pl}^2}\Big(\frac{k}{aH}\Big)^3\frac{e^{2\psi_s}}{(1-\epsilon_1)|\omega_s|}
\end{equation}
\begin{equation}
\mathcal{P}_T=\frac{16H^2}{\pi M_{pl}^2}\Big(\frac{k}{aH}\Big)^3 \frac{e^{2\psi_T}}{(1-\epsilon_1)|\omega_T|}
\end{equation}

Here we have seen  a brief on  the general cosmological perturbation equations and the application of WKB approximation on them as was given in the work 
by J. Martin and D. Schwarz (2003) \cite{mainref} and the derivations of the power spectra for the scalar and the tensor perturbations. Further we 
use the above method for Power law inflation as the model is exactly solvable and hence comparable with the observation data.

\section{Application of WKB approximation on Power law inflationary model}
For Power law inflation, the scale factor is 
\begin{equation}
a(\eta)= l_{o}|\eta|^{1+\beta}
\end{equation}
where $\beta\leq-2$ and $l_{o}$ is the Hubble's radius which is a constant when $\beta=-2$. The corresponding Horizon flow functions are given as,
\begin{equation}
\epsilon_{n+1}= \frac{d ln |\epsilon_{n}|}{dN}
\end{equation} for $n\geq0$.
 Inflation takes place when $\epsilon_{1}<1$.
 The exact power spectrum is given as 
\begin{equation}
\mathcal{P}_{\zeta}=\frac{l^{2}_{pl}}{l^{2}_{0}}\frac{f(\beta)k^{2\beta+4}}{\pi\epsilon_{1}}
\end{equation}
\begin{equation}
\mathcal{P}_{h}=\frac{l^{2}_{pl}}{l^{2}_{0}}\frac{16f(\beta)k^{2\beta+4}}{\pi}
\end{equation}
where $l_{pl}=m_{pl}^{-1}$ is the Planck Length and the function $f(\beta)$ is given by \begin{equation}
f(\beta)=\frac{1}{\pi}\Big[\frac{\Gamma(-\beta-\frac{1}{2})}{2^{\beta+1}}\Big]^{2}
\end{equation} where $\Gamma$ is the Euler integral of the second kind.

 When  $f(\beta=-2)=1$, this becomes as a special case because when $\beta=-2$ the expression of the scalar spectrum blows up.

 By applying the WKB approximation the effective frequency is given as,
 \begin{equation}
 \omega^2(\eta)=k^2-\beta(\beta+1)/\eta^2 
\label{ome}
 \end{equation}
and $|\frac{Q}{\omega^2}|=\frac{1}{4\beta(\beta+1)}$.
 As for inflation $\frac{Q}{\omega^2}\ll1$, when $\beta=-2$ the ratio goes to 1/8 hence satisfying the condition.

 By applying the WKB approximation we get the power spectra to be 
\begin{equation}
\mathcal{P}_{\zeta}=\frac{l^{2}_{pl}}{l^{2}_{0}}\frac{1}{\pi\epsilon_{1}}g(\beta)k^{2\beta+4}
\label{P1}
\end{equation}
\begin{equation}
\mathcal{P}_{h}=\frac{l^{2}_{pl}}{l^{2}_{0}}\frac{16}{\pi}g(\beta)k^{2\beta+4}
\label{P2}
\end{equation}
and the function is given as,
\begin{equation}
g(\beta)=\frac{2e^{2\beta+1}}{(2\beta+1)^{2\beta+2}} .
\end{equation}
The corresponding spectral indices are then computed to be 
\begin{equation}
n_s-1=n_T=2\beta+4 .
\end{equation}

The Constant-roll inflation or the Ultra-slow-roll inflationary model is also a viable alternative to the standard slow-roll approximation. 
In this model, the inflationary scenario is investigated with constant rate of roll, 
$\ddot{\Phi}/ H\dot{\Phi}$ = $-3-\alpha$ (remains a constant) \cite{moho}.
The model gives experimentally viable results for $\alpha \lesssim -3$. The parameter constrains on the constant
roll model and comparison with observations  is explored by Motohashi and Starobinsky  \cite{mohoo} which gives a permitted range 
of $r$ values as  $r \lesssim 0.07 $ to  $r \ll 10^{-3} $.

\section{WKB Power Spectra and Tensor to scalar ratio(\textit{r})}

By using the WKB approximation, we first see that the effective frequencies in Eq. (\ref{ome}) are given as, 
\begin{equation}
\omega_s^2=k^2\eta^2 -\Big(\frac{9}{4}+3\epsilon_1+\frac{3}{2}\epsilon
_2\Big)
\end{equation}
\begin{equation}
\omega_T^2=k^2\eta^2-\Big(\frac{9}{4}+3\epsilon_1\Big)
\end{equation}

Then corresponding WKB spectra are derived from solving  Eq. (\ref{P1}) $\& $ (\ref{P2}) with the above frequencies and we get the scalar power spectrum as,
\begin{equation}
\mathcal{P}_{\zeta}= \frac{H^{2}}{\pi\epsilon_{1}m^{2}_{pl}}(18e^{-3})[1-2(D+1)\epsilon_{1}-D\epsilon_{2}-(2\epsilon_{1}+\epsilon_{2})ln\Big(\frac{k}{k_{*}}\Big)+\textit{O}(\epsilon_{n}^{2})]
\label{pks}
\end{equation}
and the corresponding tensor power spectrum as,
\begin{equation}
\mathcal{P}_{h}=\frac{16H^{2}}{\pi m^{2}_{pl}}(18e^{-3})[1-2(D+1)\epsilon_{1}-2\epsilon_{1}ln\Big(\frac{k}{k_{*}}\Big)+\textit{O}(\epsilon_{n}^{2})]
\label{pkt}
\end{equation}
with $D\equiv\frac{1}{3}-ln3\approx-0.765$.

The Spectral indices calculated from above equations are in accordance with the spectral indices given by slow-roll inflation but it gives the next higher order
terms which increases it's accuracy. The spectral indices are obtained as,

\begin{equation}
n_{s}-1=\frac{-2\epsilon_{1}}{1-\epsilon_{1}}-\frac{\epsilon_{2}}{1-\epsilon_{1}}+
\frac{-2\epsilon_{1}-\epsilon_{2}}{[1-2(D+1)\epsilon_{1}-D\epsilon_{2}-(2\epsilon_{1}+\epsilon_{2})ln(\frac{k}{k_{*}})]}+\mathcal{O}(\dot{\epsilon_{n}})
\end{equation}

\begin{equation}
n_{T}=\frac{-2\epsilon_{1}}{1-\epsilon_{1}}+\frac{-2\epsilon_1}{[1-2(D+1)\epsilon_{1}-(2\epsilon_{1})ln(\frac{k}{k_{*}})]} + \mathcal{O}(\dot{\epsilon_{n}})
\end{equation}
With Slow-roll approximation the power spectra are \cite{stewart}
\begin{equation}
\mathcal{P}_{\zeta}=\frac{H^{2}}{\pi\epsilon_{1}m^{2}_{pl}}[1-2(C+1)\epsilon_{1}-C\epsilon_{2}-(2\epsilon_{1}+\epsilon_{2})ln\Big(\frac{k}{k_{*}}\Big)+\textit{O}(\epsilon_{n}^{2})]
\end{equation}
and the tensor powerspectrum, 
\begin{equation}
\mathcal{P}_{h}=\frac{16H^{2}}{\pi m^{2}_{pl}}[1-2(C+1)\epsilon_{1}-2\epsilon_{1}ln\Big(\frac{k}{k_{*}}\Big)+\textit{O}(\epsilon_{n}^{2})]
\end{equation}
where C=$\gamma_{E}+ln2-2\simeq$  -0.7296 , $\gamma_{E}\simeq$0.5772 which is the Euler constant. All the quantities are calculated at $\eta_*$ which is when $k_*=aH(N_*)$. This scale $k_*$ is called the Pivot scale.
The spectral indices with the slow-roll approximation are given as,
\begin{equation}
n_s-1=-2\epsilon_1-\epsilon_2, \quad n_T=-2\epsilon_1
\end{equation}
From this we could readily see that the spectral indices of the WKB approximation  reduces to the spectral indices of the slow-roll approximation.

The scalar and tensor power spectra are represented by using the Power law form. For which the expressions for Hubble's constant is 

\begin{equation}
H=-\frac{\dot{H}}{H^2}=\frac{H'}{aH^2}=(1+\beta)\eta^{-(\beta+2)}/l_o
\end{equation}
and $\epsilon_2=0$.
The power-spectra are analytically represented and by setting the pivot scale at $k_*=0.05 Mpc^{-1}$. 

The plots show the scale-invariant scalar and tensor power spectra with the slow-roll approximation and with the WKB approximation.
The power-spectra are calculated for the $\beta=-2.005$. 

\begin{figure}[!hbt]
   \centering
   \includegraphics[width=4.5in]{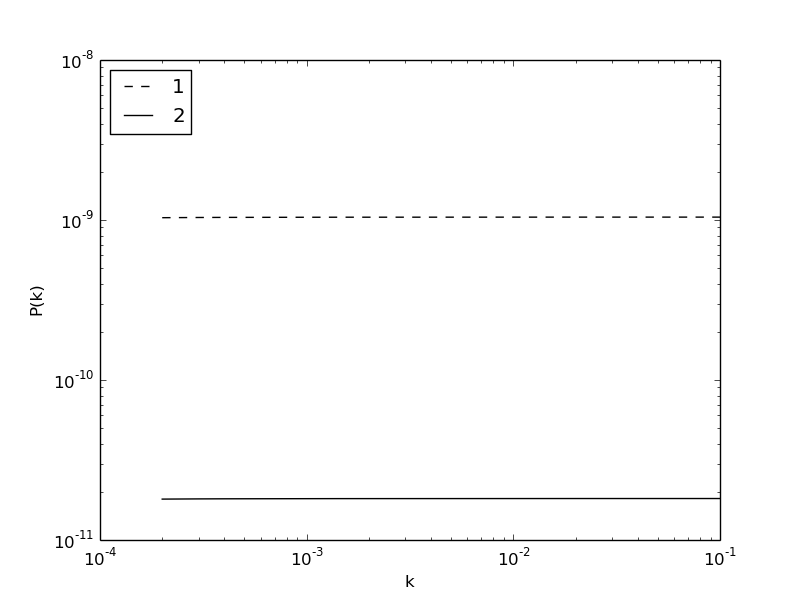}
\label{fig:1}
\caption{\small Scalar power-spectra (1)with slow-roll approximation and (2)with WKB approximation}
\end{figure}

\begin{figure}[!hbt]
    \centering
    \includegraphics[width=4.5in]{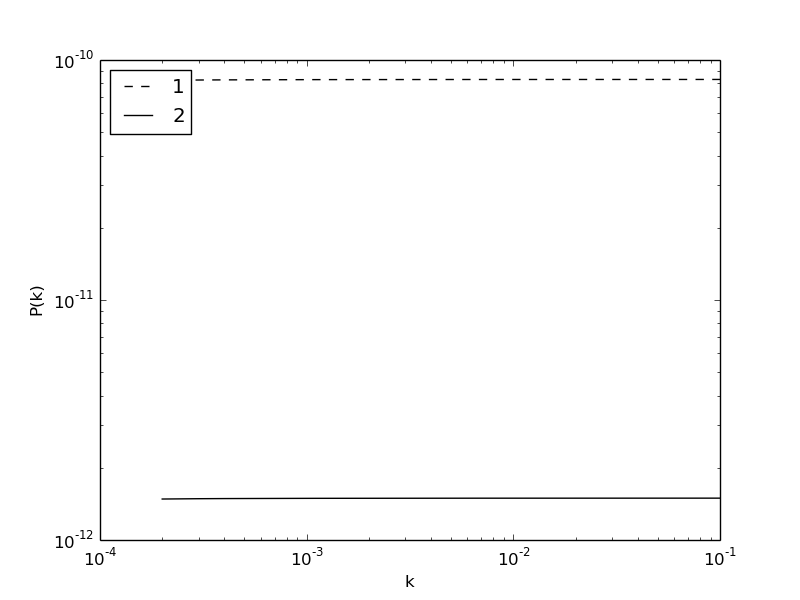}
    \caption{Tensor power spectra (1)with slow roll approximation and (2)with WKB approximation}
\label{fig:2}
\end{figure}

It is clear that the power spectra are scale invariant with both approximation techniques and the WKB approximation
 spectra shows a slight variation in the amplitude which can be compensated by changing the amplitude terms in Eq. (\ref{pks}) and (\ref{pkt}).
The corresponding spectral indices are calculated for these  spectra. With the WKB approximation it is $n_s-1=n_T=3.12177276e-04 $ and with the slow roll approximation
$ n_s-1=n_T=3.08732253e-04 $. 


The application of WKB approximation helps  in determining the power spectra even when the Slow-roll approximation is invalid and hence gives us a new frontier 
for viewing inflation. The added advantage is that it gives solutions with higher accuracy and higher order terms which is a necessity to match with the very
 precise observations of WMAP and Planck.

We use CosmoMC \cite{cosmomc}, to establish the constrains on the parameters and to identify the maximum likelihood region in the parameter space. 
The primordial power-spectrum in CAMB of CosmoMC is changed to the WKB approximation power spectrum.
As even a slight change can produce large deviations in the CMB sky, we have constrained the WKB approximation parameter 
to $\beta=-2.005$. The modified power spectrum is used to then generate the CMB angular power spectrum with flat priors and
using the whole base-set of parameters with power law $\Lambda CDM$ model.

We stay with  the minimal six parameter spatially flat  $\Lambda CDM$ cosmological model as our base model with base 
parameters as  baryon density today $\Omega_{b}h^2$, cold dark matter density today $\Omega_{c}h^2$, amplitude of scalar
 power spectrum $A_{s}$, spectral index of scalar power spectrum $n_s$, tensor-to-scalar ratio \textit{r} and the tilt of 
tensor power spectrum $n_t$ but with WKB primordial power spectrum. This model has four free non-primordial cosmological parameters same as the power-law
baseline \textit{i.e}  $\omega_b$, $\omega_c$, $\theta_{MC}$ and $\tau$. Here we try to estimate the impact of primordial power spectrum with higher-order terms of Hubble-flow
 functions, which are analogous to slow-roll parameters with emphasis on the deviations in tensor-to-scalar ratio. We also take to consideration  three neutrinos species,
 with two mass-less states and a single massive neutrino of mass $m_{\nu}$ = 0.06eV.

The Planck likelihood code$^\dag$ {\footnotetext[1]{2018 Planck CMB likelihoods (except for lensing) are not published yet.
The present discussion refer to the 2015 likelihoods.}} (PLC/clik) and parameter chains are available from the Planck Legacy Archive. 
The Planck lensing and Bicep-Keck-Planck likelihoods are included with the cosmomc. 

\begin{figure}[H]
\centering
\includegraphics[width=6.1in]{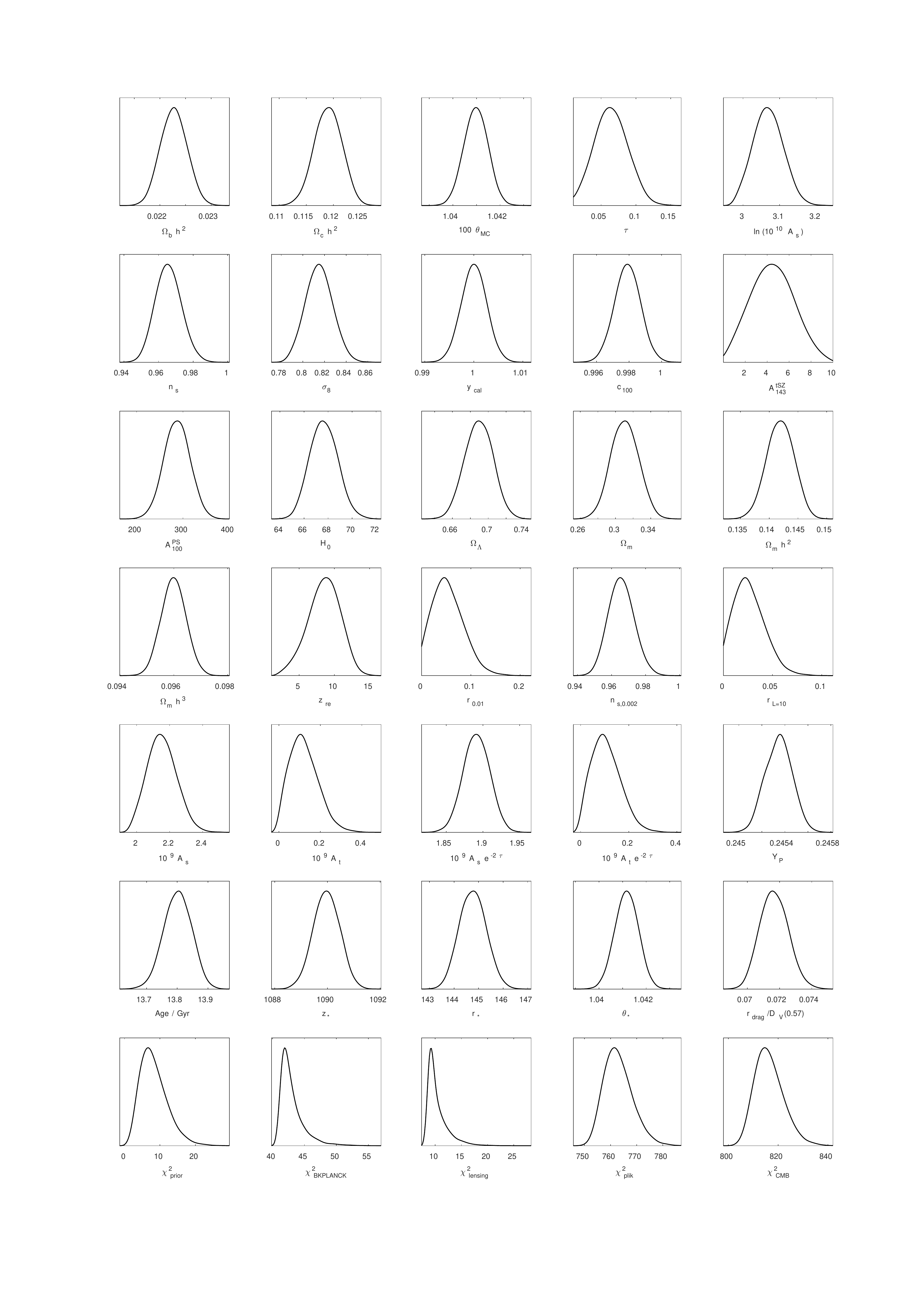}
\caption{\small 1D marginalized posterior distribution of the inflationary parameters and cosmological parameters with WKB approximation}
\label{fig:3}
\end{figure}

The limits and marginalized densities given in Fig. \ref{fig:3} are generated using the getdist package which is part of CosmoMC, it  
uses the baseline likelihood which is a hybrid, 
by connecting together a low-multipole likelihood  with the Gaussian likelihood constructed from the higher multipoles. We study the Bayesian factors
of the model
with respect to the power-law base model and the $\chi^2_{eff}$ values with power-law model. 
As it is clear the WKB power spectrum is scale invariant and the parameters are in the $95\%$ CL of Planck results.

\begin{figure}[H]
\centering
\includegraphics[width=5.1in]{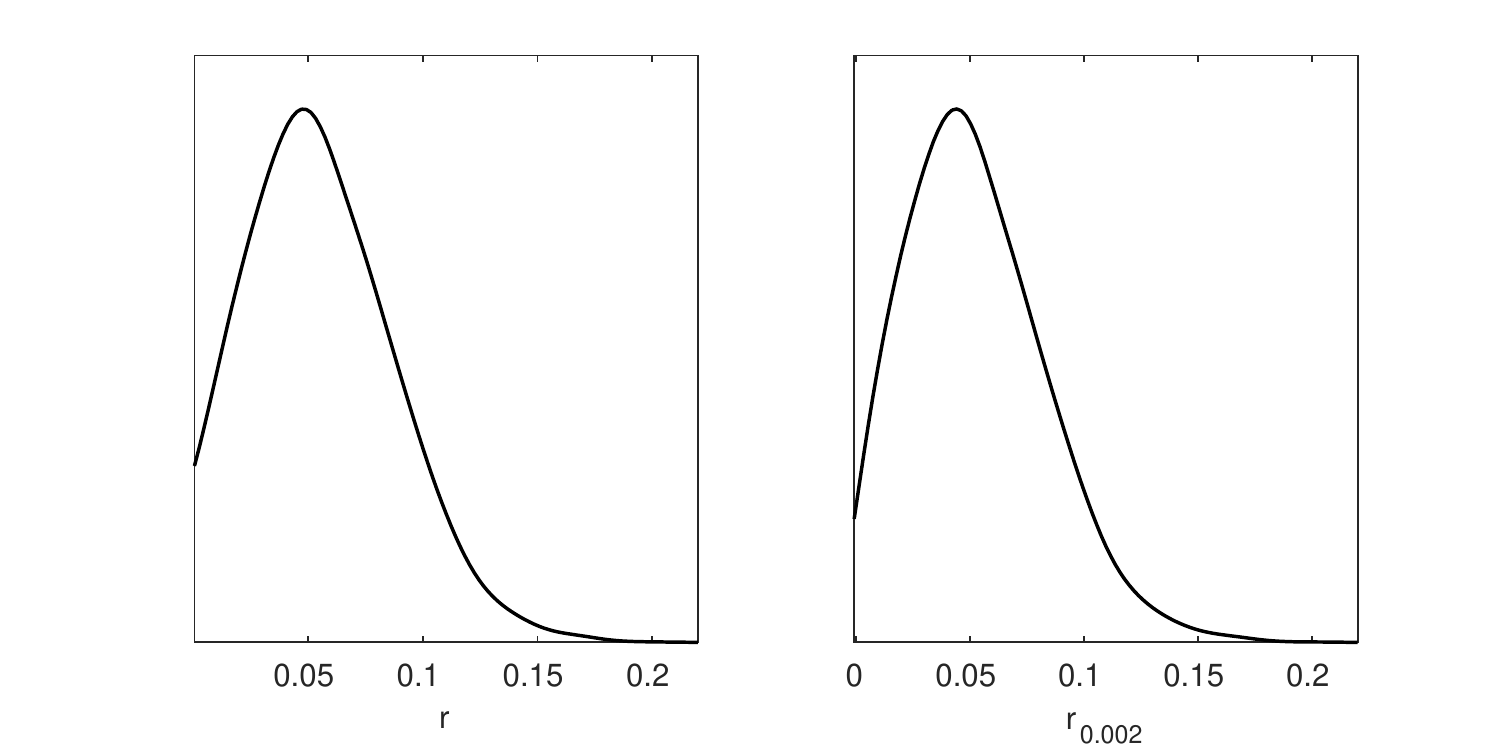}
\caption{\small 1D marginalized posterior distribution of the Tensor to Scalar ratio \textit{r} with WKB approximation}
\label{fig:4}
\end{figure}

By combining Planck temperature, low- polarization and lensing 2018 Planck results give  $r_{0.002} < 0.10 $ ($95\%$ CL, Planck TT+lowE+lensing)
\cite{planck2018}.
This constraint slightly improves on the corresponding Planck 2015, $95\%$ CL bound, i.e., $r_{0.002} < 0.11 $ \cite{planck2015} which is 
consistent with the B-mode polarization constraint $r < 0.12$ (95 $\%$ CL), obtained from a joint analysis of the BICEP2/Keck Array and Planck data. 
With the WKB approximation, the upper limit on \textit{r} is obtained 
as $\textit{r} < 0.1109$ and $\textit{r}_{0.002} < 0.1055 $ (BKPlanck$^\dag${\footnotetext[2]{BICEP2/Keck Array and 2015 Planck joint analysis}}+ lensing). 
This gives us a window into higher order corrections 
and better fit to the Planck results. It is seen that the WKB best fit value of the spectral index is $n_s=0.9657$ with $H_{o}=67.64$ which are 
in the $95\%$ CL of Planck observations.

\section{Conclusion}
Present work studied the scalar and tensor power spectra with the WKB approximation and constrained the value of the WKB parameter $\beta$. 
The WKB approximation helps in determing the power spectra even when the Slow-roll approximation is invalid, gives the higher order corrections and 
the precise details of Power law model which will give insights into paramount questions
in the Inflationary model of Universe. The spectral index values and spectra  generated with WKB approximation are compared with the observation data and we see that
 the WKB approximated power spectra could become a viable alternative for Slow-roll approximation. Running Monte-Carlo chains with CosmoMC and generating the plots with 
Getdist, the tensor to scalar ratio(\textit{r}) is also analysed by
 constraining the cosmological parameters with WKB approximation, setting an upper limit at $ \textit{r} < 0.1109 $. Here \textit{r}
is defined at the pivot scale $k_* = 0.05Mpc^{-1}$. We also report the bounds on $\textit{r}_{0.002}$, the tensor to scalar ratio at $k_* = 0.002 Mpc^{-1}$
as $\textit{r}_{0.002} < 0.1055 $ which is consistent within $1\sigma$ with the Planck results.  

\section{Acknowledgments}
We acknowledge the use of high performance computing system at  IUCAA. MJ acknowledges the support from U.G.C. Major
research project grant, $F. No.-43-523/2014(SR)$ and the Associateship of
 IUCAA.  AA acknowledges the Junior Research Fellowship from University Grants Commission (UGC), India.  We thank the referee for the valuable suggestions and 
comments. 




\end{document}